\documentclass[12pt]{article}
\usepackage{graphicx}
\usepackage{amsfonts}
\usepackage{amsmath}
\usepackage{amssymb}

\begin{document}

\title{Transformation Semigroup and Complex Topology:  a study of inversion with increasing complexity}

\author{August Lau and Chuan Yin \\
        \\
        Apache Corporation \\
        2000 Post Oak Blvd., Houston, Texas 77056 \\
        \\
        Email contact: \texttt{chuan.yin@apachecorp.com}}
\date{August 11, 2010}

\maketitle

\begin{abstract}

This paper is a continuation of our 2005 paper on complex topology and its implication on invertibility (or non-invertibility).  In this paper, we will try to classify the complexity of inversion into 3 different classes.  We will use synthetic models based on well control to illustrate the different classes. The first class is systems which have a group of symmetry.  This class has clean inversion. We will use two examples which include 2-term AVO on 1-D layers as an example.  The second class does not have a group support.  It is in general described by a semigroup which is a set of operators with or without inverses.  There is no guarantee in general of invertibility in a global sense.  Even though this class does not have invertibility in general, there could still be local weak invertibility embedded in the semigroup.  The last class is system with complex topology where the underlying topology/geometry requires infinite construction.   In our previous 2005 paper, we gave the 1-D Cantor layers as the forerunner of all complex topological interface.  We will re-examine this last class in detail.  The idea of constructing the Cantor layers as inverse limit was introduced in the previous paper.  We will examine the finite approximation of the Cantor layers and its implication on inversion.
\end{abstract}

\section*{Introduction}

We will consider seismic situations where the information is scrambled with no loss of energy. The information cannot be unscrambled in general.  We will form certain strategies to unscramble the information. 

We will restrict our discussion to 2-dimensional vector space $R \times R$ where $R$ is the set of real numbers and $R \times R$ is the plane consisting of all $(x,y)$ where $x,y$ are real numbers.  The operation $fg$ means the composition of two functions $f$ and $g$.

The concept of group can be illustrated by collecting certain functions $G = \{f | f: R \times R => R \times R ,  ff^{-1}=e\}$ where $e$ is an idempotent ($e=ee$) and $f^{-1}$ is the inverse relative to $e$.  

The concept of semigroup can be illustrated by collecting all functions $S = \{f | f : R \times R=>R \times R\}$.  

The concept of complex topology can be illustrated by collecting all compact subsets $T = \{X | X \subseteq R \times R\}$.

A concrete example for group $G$ is the set of linear functions $f(x,y)=(ax+by,ax-by)$ where $a,b$ represents coefficients and $x$ is the AVO intercept and $y$ is the AVO gradient.  Different rotation given by different $a$ and $b$ would give different estimates on lithology/fluid content.  

A concrete example for semigroup $S$ of operators is the set of all $2\times 2$ matrices.  These matrices are in general non-invertible.  The most non-invertible matrix is the zero matrix which has all 4 elements being zero.  Semigroups have parts that are invertible and parts that are non-invertible.

\section*{Class 1: Transformation Group}

A group is a set of transformations which have inverses. A common example is the set of $2\times 2$ non-singular matrices under matrix multiplication.  We will give two examples relevant to geophysics.

\subsection*{Example 1.1 : Simple multiples}

We give an example of layers which have only weak interactions.   Using this example,  we generate the interbed multiples to see what the deformation of the wavelet looks like.  The deformed wavelet has a exponential decay  curve in the amplitude spectrum.  Exponential is the simplest representation of a 1-parameter group.  It naturally has an inverse,  i.e.,  $exp(x)exp(-x) = 1$.  This is the lowest order group. The 1-D multiples form a series with varying coefficients.  We know that such a series cannot be represented by an exponent.  However, we essentially force an exponential explanation to the process so that it is easily invertible. This is a common formulation in geophysics so we can get inverses even though it is a coarse approximation.  As shown in Schoenberger and Levin paper,  the seismic response is much more complicated than an exponential. 

\begin{figure}
\centering
  \includegraphics[width=5in]{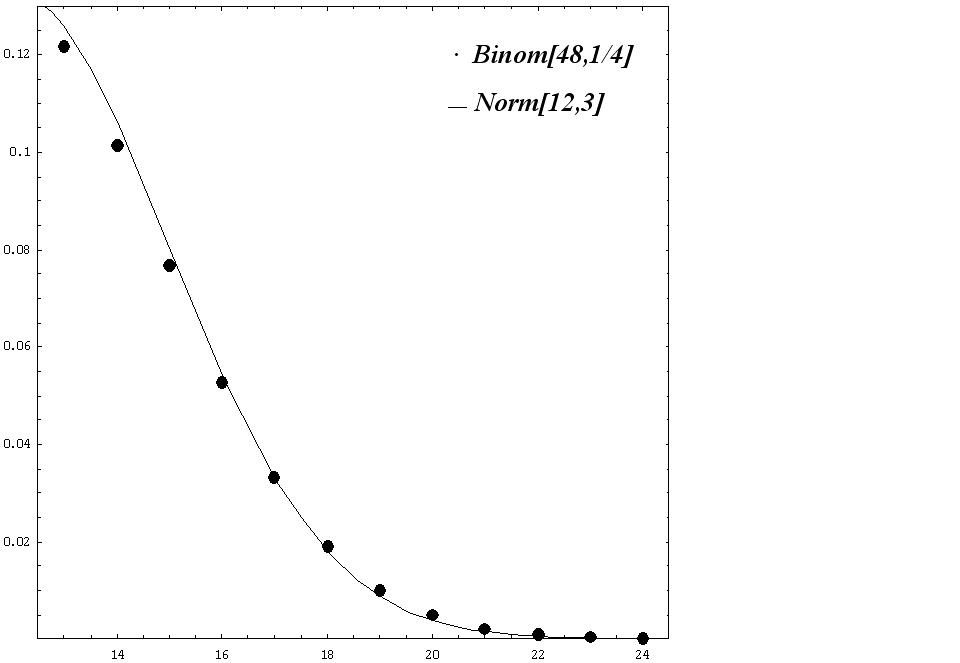}
\caption{(Example 1.1) Normal approximation to the Binomial PMF.} 
\label{fig1}
\end{figure}

\subsection*{Example 1.2 : 2-term AVO}

If we consider 1-D layers and offset dependent amplitudes in the seismic gather,  we can compress the complicated offset amplitude into intercept and gradient.  Again, it is an attempt to convert a complicated system into a group of symmetry.  The group here is the group of rotations.  If $A$ is the intercept and $B$ is the gradient,   we can rotate $A$ and $B$ into new axes.  The rotation is just the unitary group of the plane (see Lau 1980).  The rotation group gives meaning to what is lithology and what is fluid, based on the rotation matrix.  It simplifies the inversion process by making a group as the fundamental operator.

\begin{figure}
\centering
  \includegraphics[width=5in]{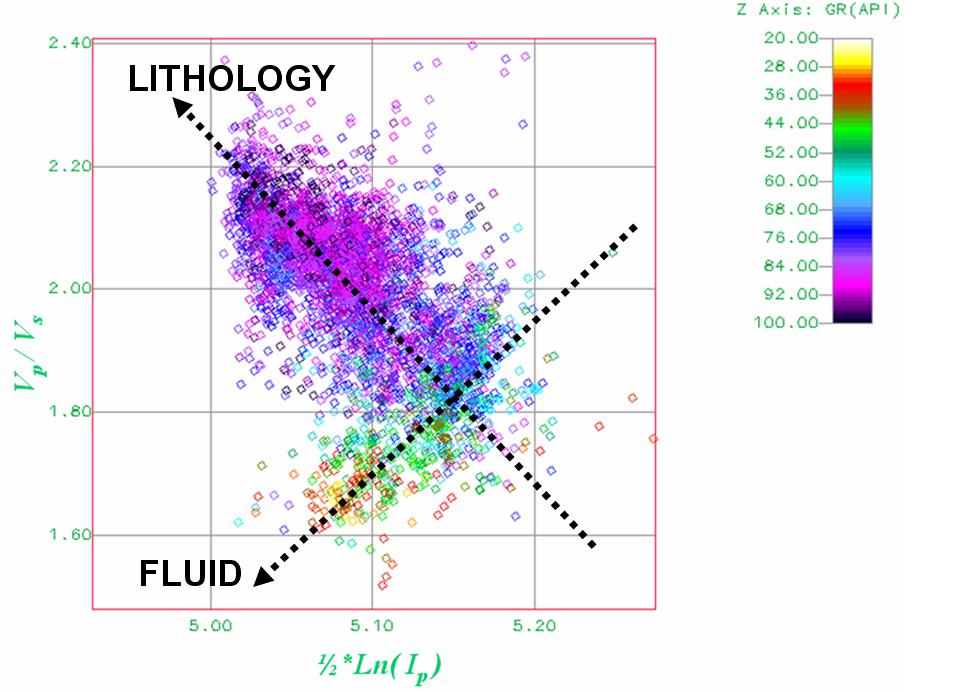}
\caption{(Example 1.2) Crossplot of acoustic impedance ($Ip$) and the $Vp/Vs$ ratio, based on wireline logs from a well in the Gulf of Mexico. Arrows suggest most sensitive directions for detecting fluid and lithology, respectively.}
\label{fig2}
\end{figure}

\section*{Class 2: Transformation Semigroup} 

Semigroup is a set of transformations which do not necessarily have inverses like group.  A common example is the set of all $2\times 2$ matrices under matrix multiplication.  The non-singular matrices do not have inverses.  We will give two examples on transformation semigroups which are relevant to geophysics.

Since semigroup is considered lossy and non-invertible,  it is viewed as undesirable.  We think that it is actually desirable to use semigroup transformations to combine terms.  If $x$ and $y$ are two variables to be inverted,   the uncertainty of $x$ and the uncertainty of $y$ can be high individually. But if we combine $x$ and $y$ into $xy$ and re-write the inversion equation,   then $xy$ is a more stable solution.   What we have done is to apply a non-invertible operator to reduce a 2-dimensional vector $(x,y)$ into a 1-dimensional vector $z$ where $z = xy$.   Of course, reduction of dimension will not be an invertible operator.  So we have transformed the inversion variables with a semigroup operator like multiplication so that the new variable has a more stable solution.  This is tantamount to reducing two large null spaces of $x$ and $y$ into a combined null space which is much smaller.  In fact ,  we do that in geophysics by combining velocity and density to form impedance which is the product of velocity and density.  So we have applied semigroup transformation in geophysics to reduce null space. Of course,  semigroup transformation is more general than just multiplication of two variables (see Hofmann and Mostert).

\subsection*{Example 2.1 : Short period multiples}

The next example is another 1-D model with zero offset synthetic.  We will insert some large reflections like air-water interface, water bottom, top salt and base salt.  The unit above top salt is generated from a complex shale unit based on a well log.  This illustrates the fact that the shale unit cannot be inverted because of the complicated short period multiples.   However, we will lump all the short period multiples into a "primary" wavelet and call it a new wavelet. Even though the problem is describable as a semigroup (non-invertible system), we reduce the non-invertible system with complex shale layering into an invertible system by lumping the short period multiples into a new wavelet  This is similar to inverse Q except with more details than approximating the apparent attenuation with exponential functions.  This is a way of forcing a non-invertible system into an invertible one by grouping the non-invertible part into a wavelet.  The challenge of this method for field data is the interpretation of what is invertible and what is not.

\begin{figure}
\centering
  \includegraphics[width=3in]{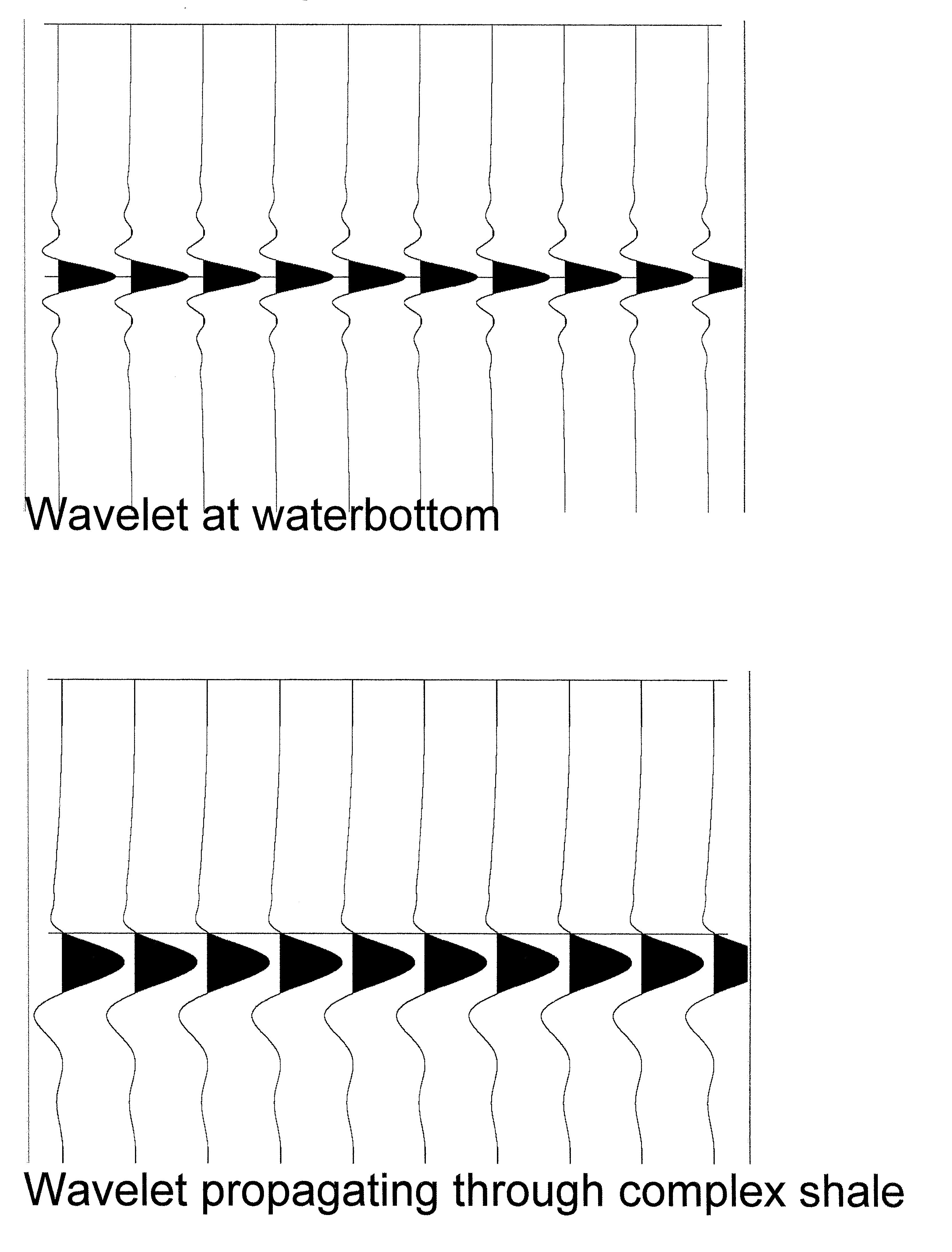}
\caption{(Example 2.1) Lump semigroup (non-invertable) part into a new wavelet.}
\label{fig3}
\end{figure}

\subsection*{Example 2.2 : Non-invertible AVO} 

If we model the AVO response precisely, it cannot be easily inverted like the rotation group used for 2-term AVO.  We will generate an example which illustrates that  the AVO is not invertible and in fact,  has cusp-like response.  So the semigroup strategy here is to approximate the exact system with quadratic terms.  Then solve the quadratic equation.   This has better behavior than the 2-term or 3-term AVO for high angles.

\begin{figure}
\centering
  \includegraphics[width=5in]{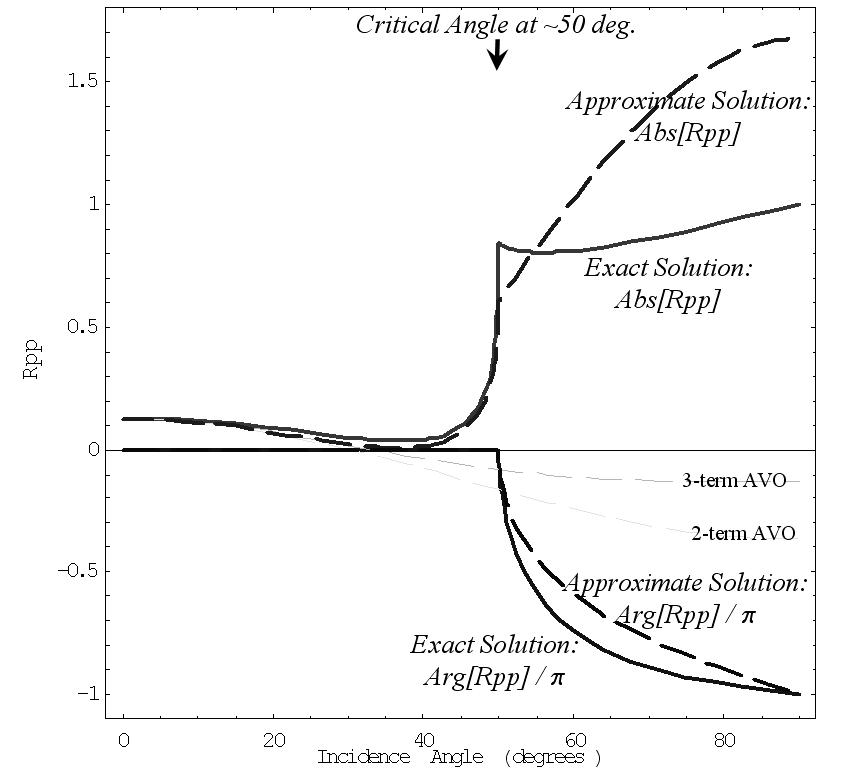}
\caption{(Example 2.2) Single-interface reflectivity for a high-impedance reservoir in the Gulf of Mexico. Solid lines are from exact solutions, and dash lines are approximate solutions. Notice the traditional 2-term and 3-term AVO approximations perform poorly beyond ~30 degrees.}
\label{fig4}
\end{figure}

\section*{Class 3: Complex Topology} 

\subsection*{Example 3.1 : Complexity of Cantor Comb (PP data)}

In our 2005 paper,  we showed the complexity of the Cantor layers which gives rise to complex multiples which are not invertible. The Cantor set is the set of real numbers which can be written as infinite sums of some power of 1/3,  the so called Cantor middle third expansion.  Its value is bounded by 0 and 1.  It has this amazing property of $C+C=I$   where $C$ is the Cantor set,  $I$ is [0,2]. We will show another rendition of the Cantor layers by looking at the Cantor comb,  i.e.,  Cantor layers standing up vertically.  This has an interesting application to fracture detection by examining the residuals.  If we perform a straightforward 1-D prestack inversion of the PP gather and then forward model the gather from the final inversion model,  we still see that the difference of the original gather and the modeled gather has a large residual.  This is due to the complex topology of the Cantor comb.

\subsection*{Example 3.2 : Complexity of Cantor Comb (PS data)}

We will generate the converted wave PS data to illustrate the complex wavefield from the Cantor comb.  Cantor comb is a toy model to understand vertical fractures.  The PS data is compared to the PP data in terms of detection.  It is shown that the residual of PS can be interpreted as vertical fractures.  The residual is again the difference between the input PS gather and the modeled PS gather from the final inversion model.
    
\begin{figure}
\centering
  \includegraphics[width=5in]{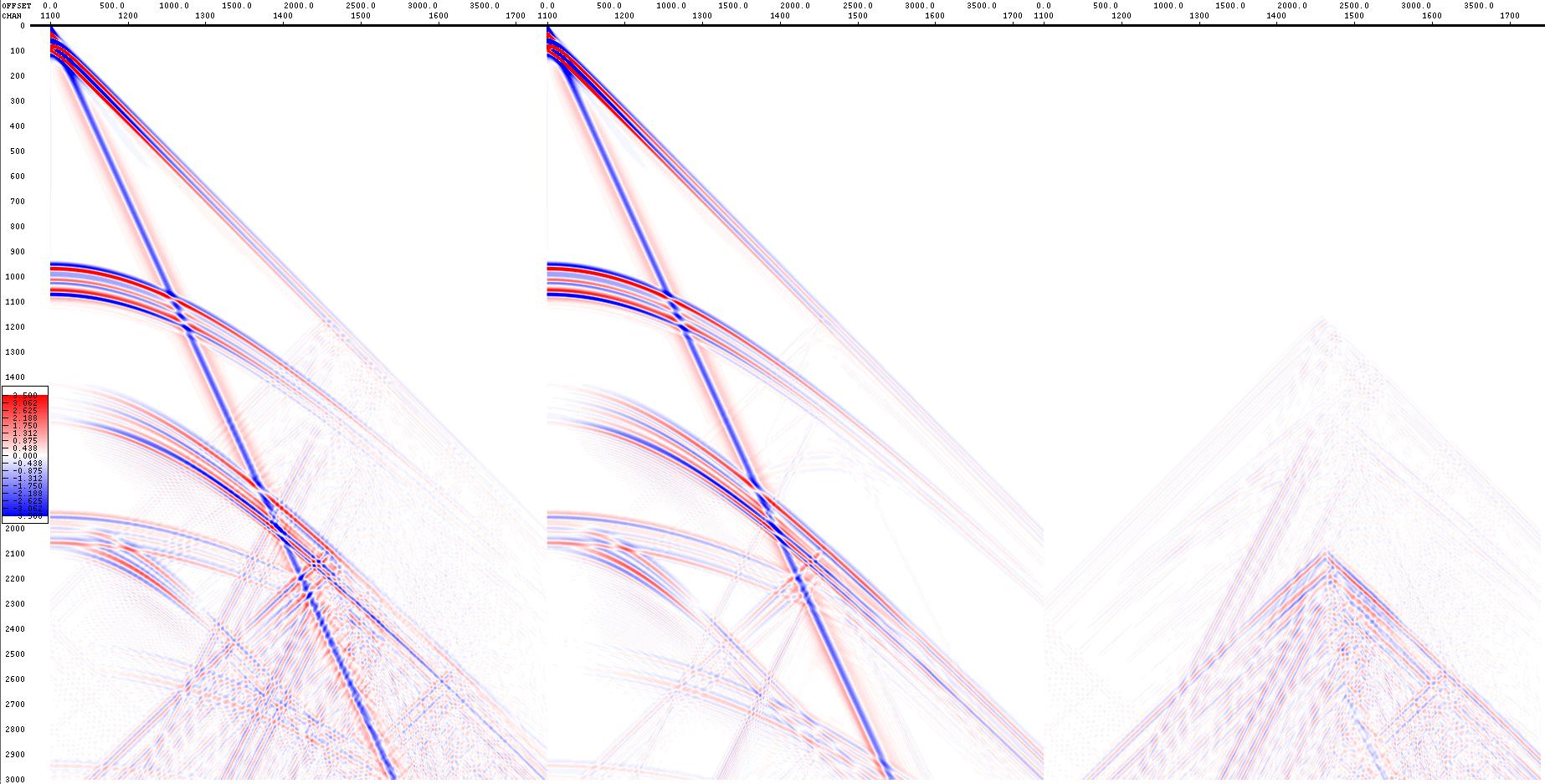}
\caption{(Examples 3.1 and 3.2) Modeled shot records from models of complex topology (left panel) and simple topology (middle panel). The right panel is the residual indicating existence of complex topology.}
\label{fig5}
\end{figure}

\subsection*{Example 3.3 : Complexity of Sierpinski Carpet}

We will generate the synthetic data to illustrate the importance of interpreting the residual.  The residual gives us an understanding of the incompleteness of the inversion and the complex geometry of the Sierpinski carpet.

\section*{Concluding remarks}

We have given three classes of examples to demonstrate increasing complexity.  The first class is clean inversion using simple group theory. This is the classical approach of rewriting systems into groups of symmetry. The second class has more complexity in the system which forms a semigroup of operators that are not invertible. We reduce the second class from semigroup to group by combining the noninvertible effects. This can then be solved using group theory. The third class is illustrated by infinitely complex system like Cantor set. This class can be interpreted using the residuals.  We hope to stimulate further discussion as to how to interpret the results when conventional inversion fails in these synthetic examples.  We suggest that we still should do the inversion.  But then we should also interpret the residual after the final inversion is done.

We have come to the meta-equation:

\begin{equation*}
Data = Class 1 + Class 2 + Class 3.
\end{equation*}

Class 1 is invertible and class 2 is not numerically invertible (lump noninvertible parts) and class 3 is due to complex topology/geometry which can be studied through the residual after inversion.  The residuals show us the discontinuities of the system even after our best effort in inversion.

In the metaphor of complex decomposition, the simple part of the data is explainable by transformation groups or semigroups (differential equations). The complex part of the data ("residual") is interpretable but not explainable by numerical methods.

\begin{figure}
\centering
  \includegraphics[width=6in]{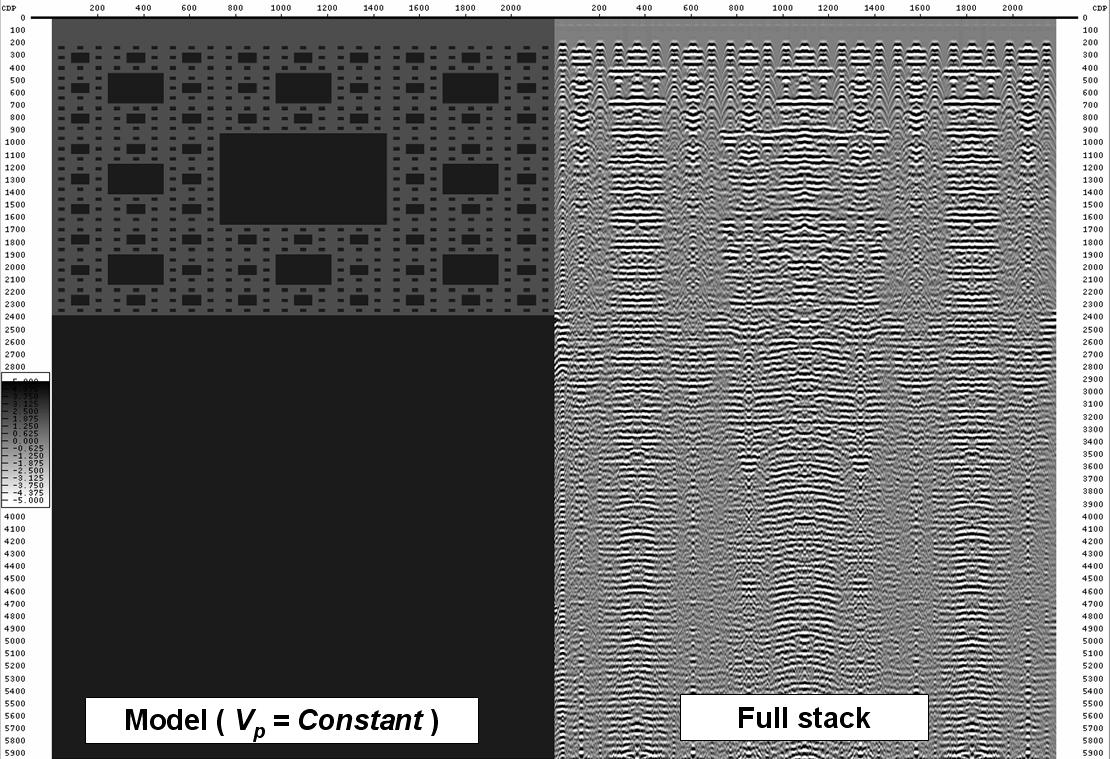}
\caption{(Examples 3.3) Sierpinski carpet creates complex multiples.}
\label{fig6}
\end{figure}

\section*{Appendix}

Figure 6 (Example 3.2):  Sierpinski carpet creates complex multiples.
Another way to improve non-invertibility is to introduce more terms in the modeling equation.   This sounds contradictory.  Let us examine some equations.

The classical wave equation can be written formally as a second order partial differential equation (pde) :
\begin{equation*}
D_{xx} = D_{tt}.
\end{equation*}

The classical heat equation can be written formally as a mixed first and second order pde:
\begin{equation*}
D_{xx} = D_t.
\end{equation*}

The heat equation can be thought in a very general way as a heat term which might or might not have much to do with heat in nature.   But it has a lot to do with lossy medium.  We propose a heat-wave equation which is highly non-linear and non-invertible.  We write the formal equation by dropping all coefficients which can be constants or functions.  The heat-wave equation is :
\begin{equation*}
D_{xx} +D_x +D_t +D_{xt} =D_{tt}.
\end{equation*}

By introducing new terms in the wave equation,  we have absorbed or lumped the non-invertible part into a new derivative or derivatives. 

\bibliographystyle{plain}
\bibliography{semigroup}   

\begin{thebibliography}{1}

\bibitem{fy}
Foster, D. J., and Yin, C., 1995, Wave propagation in elastic thin layers, Mathematical Methods in Geophysical Imaging III, Proc. SPIE V.2571, The International Society for Optical Engineering.

\bibitem{goup}
Goupillaud, P. L., 1992, Three new mathematical developments on the geophysical exploration horizon,The Leading Edge June 1992, 40-42.

\bibitem{hock}
Hocking, J. G., and G. S. Young., 1961, Topology, Addison-Wesley.

\bibitem{hof}
Hofmann, K. H. and P. S. Mostert,  1966,  Elements of Compact Semigroups,  Merrill Books Inc.

\bibitem{lau}
Lau, A. , 1980, Plane continua and transformation groups, Proceedings of the American Mathematical Society, No 4 Vol 78,  68-610.

\bibitem{ly}
Lau, A. and C. Yin, 2005, Complex topology: its impact on seismic inversion and modeling in single component recording and multicomponent recording, Internat. Mtg of Society of Exploration Geophysicists, Expanded Abstracts 1666-1668, Houston, Texas.

\bibitem{ly4}
Lau, A. and C. Yin, 2009, Geometric simplicity as a migration criterion: an application of computational topology to seismic imaging, Internat. Mtg of Society of Exploration Geophysicists, Expanded Abstracts 2773-2777, Houston, Texas.

\bibitem{ly2}
Lau, A. and C. Yin, 2010, L0+L1+L2 mixed optimization:  a geometric approach to seismic imaging and inversion using concepts in topology and semigroup, arXiv:1007.1880v1.


\bibitem{ly3}
Lau, A., C. Yin, R. R. Coifman and A. Vassiliou, 2009, Diffusion semigroups: a diffusion-map approach to nonlinear decomposition of seismic data without predetermined basis, Internat. Mtg of Society of Exploration Geophysicists, Expanded Abstracts 2327-2331, Houston, Texas.

\bibitem{sl}
Schoenberger, M. and F. K. Levin,  1978, Apparent attenuation due to intrabed multiples, II, Geophysics, Vol. 43, No. 4, 1978, 730-737.

\bibitem{yama}
Yamaguti, M.,  Hata, M., and Kigami J., 1997,  Mathematics of fractals, Translations of Mathematical Monographs,  American Mathematical Society.

\end{thebibliography}

\end{document}